\documentclass[12pt,thmsa]{article}
\usepackage{amssymb}
\usepackage{sw20lart}

\input{tcilatex}

\begin{document}

\author{{\large F. Iddir}$\thanks{
E-mail: \emph{iddir@univ-oran.dz}}${\large \ }and {\large \ L. Semlala}$%
\thanks{
E-mail: \emph{l\_semlala@yahoo.fr}}$ \and $Laboratoire${\small \ }$de$%
{\small \ }$Physique${\small \ }$Th\acute{e}orique,${\small \ }$Univ.$%
{\small \ }$d^{\prime }Oran${\small \ }$Es${\small -}$S\acute{e}nia${\small %
\ }$31100$ \and $ALGERIA$}
\title{{\LARGE Hybrid Mesons Masses in a Quark-Gluon Constituent Model}}
\date{(Third version) \\
24 December 2004 }
\maketitle

\begin{abstract}
QCD theory allows the existence of states which cannot be built by the na%
\"{\i}ve quark model; both theoretical arguments and experimental data
confirm the hypothesis that gluons may have freedom degrees at the
constituent level, and should be confined. In this work, we use a
phenomenological potential motivated by QCD (with some relativistic
corrections) to determine the masses and the wavefunctions of several hybrid
mesons, within the context of a constituent \textit{q\={q}g }model. We
compare our estimates of the masses with the predictions of other
theoretical models and with the observed masses of candidates.
\end{abstract}

\section{Introduction}

Quantum Chromodynamics, acknowledged as the theory of strong interactions,
allows that mesons containing gluons (as \emph{q\={q}g} hybrids) may exist.
The physical existence of these ``\emph{exotic}'' particles (beyond the
quark model) is one of the objectives of experimental projects $^{[1]}$.
These programs would contribute significantly to the future investigation of
QCD exotics, and should improve our understanding of hybrid physics and on
the role of the gluon in QCD. From experimental efforts at IHEP $^{[2]}$,
KEK $^{[3]}$, CERN $^{[4]}$ and BNL $^{[5]}$, several hybrid candidates have
been identified, essentially with exotic quantum numbers $J^{PC}=1^{-\text{ }%
+}$.

Hybrids have been studied, using the Flux-Tube Model $^{[6]}$, the Quark
Model with a Constituent Gluon $^{[7-9]}$, the MIT Bag Model $^{[10]}$, the
Lattice Gauge Theory $^{[11]}$, QCD Sum Rules $^{[12]}$ and using Many-Body
Coulomb Gauge Hamiltonian $^{[13]}$. These models predicted that the
lightest hybrid meson will be the $J^{PC}=1^{-\text{ }+}$ meson, in \emph{%
1.5-2.1 GeV }mass range; a charmed hybrid meson will have a mass around $4.0$
$GeV$, and a bottom hybrid meson with a mass around $10.0$ $GeV$.

We propose estimates for the masses of the hybrid mesons considering the
five $(u,$ $d,$ $s,$ $c$ and $b)$ flavors, within the context of a
Quark-Gluon Constituent Model using a QCD-inspired potential and taking into
account some relativistic effects.

\section{The phenomenological potential model}

Although the Lagrangian of QCD is known, its theory is not able to describe
unambiguously the strong-coupling regime. In this framework, the process
requests alternative theories like Flux-Tube model, Bag model, QCD string
model, Lattice QCD, or phenomenological potential models.

The potential model is essentially motivated by the experiment, and its wave
functions are used to represent the states of the strong interaction and to
describe the hadrons. The most used is the harmonic oscillator potential,
which gives very simple calculations and which is qualitatively in good
agreement with the experimental data; although its expression is not
explicitly describing the QCD characteristics, namely the (linear)
confinement and the asymptotic freedom.

The most usual kinds of potential models are using non relativistic
kinematics, which is convenient to the heavy flavors systems, but cannot be
suitable to the hadrons containing light flavors. In this work, we consider
relativistic systems to adjust the situation and then extend the study to
such bound states.

We introduce a model in which the gluon is considered as massive constituent
particle. The constituent gluon mass ($m_{g}\simeq 800$ $MeV$) is choosen as
an order of magnitude, taking in account the mass of the glueball candidate (%
$1.6$ $GeV$). Furthermore, the authors of [$21$] are generating constituent
quarks and gluon masses in the context of Dynamical Quark Model employing
BCS vacuum, and obtained a constituent gluon mass of $800$ $MeV$. As we will
see below, the impact of the $m_{g}$ in the results of hybrid masses is weak
and the order of magnitude remains the same.

The Hamiltonian is constructed, containing a phenomenological potential
which reproduce the QCD characteristics; its expression has the mathematical
``Coulomb + Linear'' form, and we take into account also some additional
spin effects.

The basic hypothesis is to use a relativistic Schr\"{o}dinger-type wave
equation$^{[14]}$: 
\begin{equation}
\left\{ \sum\limits_{i=1}^{N}\sqrt{\vec{p}_{i}^{\text{ }2}+m_{i}^{\text{ }2}}%
+V_{eff}\right\} \text{ }\Psi (\vec{r}_{i})\text{ }=E\text{ }\Psi (\vec{r}%
_{i}).  \tag{1}
\end{equation}

Another wave equation, more convenient for multiparticle systems, can be used%
$^{[15,\text{ }16]}$: 
\begin{equation}
\left\{ \sum\limits_{i=1}^{N}\left( \frac{\vec{p}_{i}^{\text{ }2}}{2M_{i}}+%
\frac{M_{i}}{2}+\frac{m_{i}^{2}}{2M_{i}}\right) +V_{eff}\right\} \text{ }%
\Psi (\vec{r}_{i})\text{ }=E\text{ }\Psi (\vec{r}_{i})\text{ };  \tag{2}
\end{equation}%
where $M_{i}$ are some \textquotedblleft dynamical masses\textquotedblright\
satisfying the conditions: 
\begin{equation}
\frac{\partial E}{\partial M_{i}}=0\text{ };  \tag{3}
\end{equation}%
$V_{eff}$ is the average over the color space of chromo-spatial potential$%
^{[17]}$: 
\begin{eqnarray}
V_{eff} &=&\left\langle V\right\rangle _{color}=\left\langle
-\sum\limits_{i<j=1}^{N}\mathbf{F}_{i}\cdot \mathbf{F}_{j}\text{ }%
v(r_{ij})\right\rangle _{color}  \nonumber \\
&=&\sum\limits_{i<j=1}^{N}\alpha _{ij}v(r_{ij})\text{ };  \TCItag{4}
\end{eqnarray}%
where $v(r_{ij})$ is the phenomenological potential term.

For the hybrid meson, $i$ and $j$ representing the constituants:

\begin{center}
$i,j=1\equiv q$

$i,j=2\equiv \overline{q}$

$i,j=3\equiv g.$
\end{center}

We consider all the possible $\overrightarrow{S_{i}}.\overrightarrow{S_{j}}$
combinations ($\alpha _{q\overline{q}}=-\frac{1}{6}\ ;\ \alpha _{qg}=\alpha
_{\overline{q}g}=\frac{3}{2}$).

We have chosen a QCD-motivated potential which has the form: 
\begin{equation}
v(r_{ij})=-\frac{\alpha _{s}}{r_{ij}}+\sigma \text{ }r_{ij}+c\text{ }; 
\tag{5}
\end{equation}
the $\alpha _{s}$, $\sigma $, and $c$ may be fitted by experimental data or
taked from lattice and Regge fits.

Whereas, for light quarks we should add the spin-dependent correction
represented by the (smeared) hyperfine term of Breit-Fermi interaction$^{[18,%
\text{ }19]}$: 
\begin{equation}
V_{S}=\sum_{i<j=1}^{N}\alpha _{ij}\frac{8\pi \alpha _{h}}{3M_{i}M_{j}}\frac{%
\sigma _{h}^{3}}{\sqrt{\pi ^{3}}}\exp (-\sigma _{h}^{2}\text{ }r_{ij}^{2})%
\text{ }\mathbf{S}_{i}\cdot \mathbf{S}_{j};  \tag{6}
\end{equation}%
we neglect the tensor and spin-orbit terms which effects are supposed to be
small$^{[14,\text{ }19]}$.

\section{The hybrid mesons}

\subsection{The quantum numbers}

For the classification of hybrid mesons in a constituent model we will use
the notations of $[8]:$

\emph{l}$_{\text{g }}$\emph{\ \ }: is the relative orbital momentum of the
gluon in the \emph{q\={q}} center of mass;

\emph{l}$_{\text{\textit{q\={q}}}}$\emph{\ \ }: is the relative orbital
momentum between \emph{q} and \emph{\={q}};

\emph{S}$_{\text{\textit{q\={q}}}}$\emph{\ }: is the total quark spin;

\emph{j}$_{\text{\textit{g }}}$\emph{\ \ }: is the total gluon angular
momentum;

\emph{L \ \ }: \textit{l}$_{\text{\textit{q\={q}}}}$ + \textit{j}$_{\text{%
\textit{g}}}.$

Considering the gluon moving in the framework of the q$\overline{q}$ pair,
the parity of the hybrid will be:

\begin{center}
$P=P\left( q\overline{q}\right) .P\left( g\right) .P\left( relative\right) $

$=\left( -\right) ^{l_{q\overline{q}}+1}.\left( -1\right) .\left( -\right)
^{l_{g}};$
\end{center}

$\left( -1\right) $ being the intrinsic parity of the gluon. Then the parity
of hybrid meson will be:

\begin{center}
$P=\left( -\right) ^{l_{q\bar{q}}+l_{g}}.$
\end{center}

The charge conjugation is given by: 
\[
C=\left( -\right) ^{l_{q\bar{q}}+S_{q\bar{q}}+1}. 
\]

Some $J^{PC}$ states are given in Table 1 where lower sign stands for the
state with \textit{S}$_{\text{\textit{q\={q}}}}=0$. Our attention is taking
aim essentially to the $1^{-\text{ }+}$ states, which are predicted to be
the lightest hybrid states by the theoretical models; furthermore some $1^{-%
\text{ }+}$ candidates have been observed.

Let us consider the lightest $1^{-\text{ }+}$ hybrid mesons: \textit{S}$_{%
\text{\textit{q\={q}}}}=0$, \textit{l}$_{\text{\textit{q\={q}}}}=1$ and 
\textit{l}$_{\text{\textit{g}}}=0,$ which we shall refer as \emph{the
quark-excited hybrid (}QE\emph{)}, and \textit{S}$_{\text{\textit{q\={q}}}%
}=1 $, \textit{l}$_{\text{\textit{q\={q}}}}=0$ and \textit{l}$_{\text{%
\textit{g}}}=1,$ which we shall refer as \emph{the gluon-excited hybrid (}GE%
\emph{).}

\subsection{The Hamiltonian and the wavefunctions}

We have to solve the wave equation relative to the Hamiltonian: 
\begin{equation}
H=\sum\limits_{i=q,\text{ }\bar{q},\text{ }g}\left( \frac{\vec{p}_{i}^{\text{
}2}}{2M_{i}}+\frac{M_{i}}{2}+\frac{m_{i}^{2}}{2M_{i}}\right) +V_{eff\text{ }%
};  \tag{7}
\end{equation}
with, for the hybrid meson (4): 
\[
\begin{array}{l}
\alpha _{q\bar{q}}=-\frac{1}{6}; \\ 
\alpha _{\bar{q}g}=\alpha _{qg}=\frac{3}{2}.%
\end{array}
\]

We define the Jacobi coordinates: 
\[
\begin{array}{l}
\vec{\rho}=\vec{r}_{\bar{q}}-\vec{r}_{q}; \\ 
\vec{\lambda}=\vec{r}_{g}-\frac{M_{q}\vec{r}_{q}+M_{\bar{q}}\vec{r}_{\bar{q}}%
}{M_{q}+M_{\bar{q}}}.%
\end{array}
\]

Then, the relative Hamiltonian is given by: 
\begin{equation}
H_{R}=\frac{\vec{p}_{\rho }^{2}}{2\mu _{\rho }}+\frac{\vec{p}_{\lambda }^{2}%
}{2\mu _{\lambda }}+V_{eff}(\vec{\rho},\vec{\lambda})+\frac{M_{q}}{2}+\frac{%
m_{q}^{2}}{2M_{q}}+\frac{M_{\bar{q}}}{2}+\frac{m_{\bar{q}}^{2}}{2M_{\bar{q}}}%
+\frac{M_{g}}{2}+\frac{m_{g}^{2}}{2M_{g}};  \tag{8}
\end{equation}
with 
\[
\begin{array}{l}
\mu _{\rho }=\left( \frac{1}{M_{q}}+\frac{1}{\text{ }M_{\bar{q}}}\right)
^{-1} \\ 
\mu _{\lambda }=\left( \frac{1}{M_{g}}+\frac{1}{M_{q}+M_{\bar{q}}}\right)
^{-1};%
\end{array}
\]
and 
\begin{eqnarray}
V_{eff}(\vec{\rho},\vec{\lambda}) &=&-\alpha _{s}\left( -\frac{1}{6\rho }+%
\frac{3}{2}\frac{1}{\left| \vec{\lambda}+\frac{\vec{\rho}}{2}\right| }+\frac{%
3}{2}\frac{1}{\left| \vec{\lambda}-\frac{\vec{\rho}}{2}\right| }\right)
+\sigma \left( -\frac{1}{6}\rho +\frac{3}{2}\left| \vec{\lambda}+\frac{\vec{%
\rho}}{2}\right| +\frac{3}{2}\left| \vec{\lambda}-\frac{\vec{\rho}}{2}%
\right| \right) +  \nonumber \\
&&+\frac{17}{6}c+V_{S}\text{ }.  \TCItag{9}
\end{eqnarray}

We have chosen to develop the spatial wave function as follows: 
\begin{equation}
\psi ^{l_{q\bar{q}}l_{g}}(\vec{\rho},\vec{\lambda})=\sum%
\limits_{n=1}^{N}a_{n}\varphi _{n}^{l_{q\bar{q}}l_{g}}(\vec{\rho},\vec{%
\lambda})\text{ };  \tag{10}
\end{equation}
where $\varphi _{n}^{l_{q\bar{q}}l_{g}}(\vec{\rho},\vec{\lambda})$ are the
Gaussian-type functions: 
\begin{equation}
\varphi _{n}^{l_{q\bar{q}}l_{g}}(\vec{\rho},\vec{\lambda})=\rho ^{l_{q\bar{q}%
}}\lambda ^{l_{g}}\exp \left( -\frac{1}{2}n\text{ }\beta _{N}^{2}\text{ }%
\left( \rho ^{2}+\lambda ^{2}\right) \right) \mathbf{Y}_{l_{q\bar{q}}m_{q%
\bar{q}}}(\Omega _{\rho })\mathbf{Y}_{l_{g}m_{g}}(\Omega _{\lambda }). 
\tag{11}
\end{equation}

Thus we solve the eigenvalue problem: 
\[
H_{nm}a_{m}=\epsilon _{N}\text{ }N_{nm}a_{m}\text{ }; 
\]
where:

\[
H_{nm}=\int d\vec{\rho}\text{ }d\vec{\lambda}\text{ }\varphi _{n}^{l_{q\bar{q%
}}l_{g}}(\vec{\rho},\vec{\lambda})^{*}H_{R\text{ }}\varphi _{m}^{l_{q\bar{q}%
}l_{g}}(\vec{\rho},\vec{\lambda}) 
\]
\[
N_{nm}=\int d\vec{\rho}\text{ }d\vec{\lambda}\text{ }\varphi _{n}^{l_{q\bar{q%
}}l_{g}}(\vec{\rho},\vec{\lambda})^{*}\varphi _{m}^{l_{q\bar{q}}l_{g}}(\vec{%
\rho},\vec{\lambda})\text{.} 
\]

This process yields $\epsilon _{N}(\beta _{N},$ $M_{q},$ $M_{\bar{q}},$ $%
M_{g})$ which is then minimized with respect to parameters $\beta _{N},$ $%
M_{q},$ $M_{\bar{q}}$ and $M_{g}.$

For the potential parameters, we must distinguish between the light and the
heavy flavors.

In Table 2 we present the parameters, fitting to the low lying isovector S-,
P- and D-wave states of the light meson spectrum (except the mass of the
strange quark obtained by fitting to the mass of Kaons)$^{[19]}.$

In Table 3 we give the parameters fitting to $J^{PC}=1^{-\text{ }-}$ \textit{%
(c\={c})}\ and \textit{(b\={b})}\ spectrum.

\section{Hybrid mesons masses}

We present in Table 4 our estimates of hybrid mesons masses for different
flavors without spin effects; we take $800$ $MeV$ for the mass of the gluon.
We compare our results with the predictions by other models.

We find the masses of the hybrid mesons larger in the \emph{GE} mode than
the \emph{QE} mode. Indeed, the strong force being proportional to the color
charge, the exchange of a color octet does require an important energy. Then
the GE hybrid meson will be heavier than the QE hybrid, which is lower
attractive.

We note that the mass of the QE hybrid state exceeds the mass of the
corresponding orbitally excited $q\overline{q}$ meson state (which is of
mass around $\sim 1.2\ GeV$ like f$_{1}$ or b$_{1}$) by $\sim 0.1\ GeV$, the
mass of the added gluon being $0.8\ GeV$ . But we cannot consider the $q%
\overline{q}$ system of the $q\overline{q}g$ hybrid meson like the
corresponding meson of mass $1.2\ GeV$: the $q\overline{q}$ system of the
excited meson being a color singlet, and the $q\overline{q}$ system of the
hybrid meson being a color octet, the difference has an important impact on
the interaction between $q$ and $\overline{q}$; then the $\alpha _{ij}$
color coefficients (equ.4) will be different, and we have:

\begin{center}
$\frac{V_{\left( q\overline{q}\right) _{1}}}{V_{\left( q\overline{q}\right)
_{8}}}=-8.$
\end{center}

For the orders of magnitude of the masses, our results are in good agreement
with the masses obtained by QCD Sum Rules$^{[12]}$, Lattice QCD $^{[11]}$,
Flux-Tube Model$^{[6]}$, Bag Model$^{[10]}$, (massless) Const. Gluon Model$%
^{[9]}$ and with the observed masses of candidates (namely $J^{PC}=1^{-\text{
}+}$ at $1.4$ and $1.6$ $GeV$\ ).

As menshioned above to check the impact of the gluon mass value on the
results, we have calculated hybrid mesons masses taking different values of $%
m_{g}$ (Table 5): the order of magnitude of the hybrid mass remains the same.

\section{The mixed-hybrid states and spin effects}

The followed expansion representing the hybrid wave function in the cluster
approximation: 
\begin{eqnarray*}
\Psi _{JM}(\vec{\rho},\vec{\lambda}) &=&\sum\limits_{n,\text{ }l_{q\bar{q}},%
\text{ }l_{g}}a_{n}^{l_{q\bar{q}}l_{g}}\sum\limits_{j_{g},\text{ }L,\text{ (}%
m),\text{ (}\mu )}\varphi _{n}^{l_{q\bar{q}}l_{g}}(\vec{\rho},\vec{\lambda})%
\text{\textbf{e}}^{\mu _{g}}\chi _{_{S_{_{q\bar{q}}}}}^{^{\mu _{q\bar{q}%
}}}\left\langle l_{g}m_{g}1\mu _{g}\mid j_{g}M_{g}\right\rangle \\
&&\times \left\langle l_{q\bar{q}}m_{q\bar{q}}j_{g}M_{g}\mid Lm\right\rangle
\left\langle LmS_{q\bar{q}}\mu _{q\bar{q}}\mid JM\right\rangle \text{ .}
\end{eqnarray*}

For the $J^{PC}=1^{-\text{ }+}$ states, restricting ourselves to the first
orbital excitations ( $l_{q\bar{q}}$ and $l_{g}\leqslant 1$\ ) we can expand
a mixing of the two modes (QE and GE): 
\[
\Psi _{1^{-\text{ }+}}(\vec{\rho},\vec{\lambda})\simeq
\sum\limits_{n=1}^{N}a_{n}^{QE}\varphi _{n}^{QE}(\vec{\rho},\vec{\lambda}%
)+\sum\limits_{n=1}^{N}a_{n}^{GE}\varphi _{n}^{GE}(\vec{\rho},\vec{\lambda}%
). 
\]

For the spin states we choused $\left\{ \left| S_{q\bar{q}},\text{ }s_{g};%
\text{ }S\right\rangle \right\} $ ( $s_{g}=1$ et $\mathbf{S}=\mathbf{S}_{q%
\bar{q}}+\mathbf{s}_{g}$ ).\linebreak

\underline{\emph{Numerical results}}

Our numerical results show that the QE-hybrid and the GE-hybrid mix very
weakly: 
\[
\begin{array}{ll}
\left| 1^{-\text{ }+}(u\bar{u}g)\right\rangle \simeq -.999\left|
QE\right\rangle +.040\left| GE\right\rangle ; & E\simeq 1.34\text{ }GeV \\ 
\left| 1^{-\text{ }+}(u\bar{u}g)\right\rangle \simeq -\left| GE\right\rangle
; & E\simeq 1.72\text{ }GeV%
\end{array}
\]

\[
\begin{array}{ll}
\left| 1^{-\text{ }+}(s\bar{s}g)\right\rangle \simeq -.999\left|
QE\right\rangle +.050\left| GE\right\rangle ; & E\simeq 1.60\text{ }GeV \\ 
\left| 1^{-\text{ }+}(s\bar{s}g)\right\rangle \simeq -\left| GE\right\rangle
; & E\simeq 2.02\text{ }GeV%
\end{array}
\]

\[
\begin{array}{ll}
\left| 1^{-\text{ }+}(c\bar{c}g)\right\rangle \simeq -.999\left|
QE\right\rangle -.040\left| GE\right\rangle ; & E\simeq 4.10\text{ }GeV \\ 
\left| 1^{-\text{ }+}(c\bar{c}g)\right\rangle \simeq -.031\left|
QE\right\rangle -.999\left| GE\right\rangle ; & E\simeq 4.45\text{ }GeV%
\end{array}
\]
\linebreak

In fact these results are not altered by the spin-spin interaction and then
are true for a $J^{PC}=1^{-\text{ }-}$.

In Table 6 we present the $1^{-\text{ }+}$ light hybrid mesons masses
calculated within spin-spin corrections (6). Note that for the string
parameter $\sigma =\frac{3}{4}0.20$ the mass of the lightest $1^{-\text{ }+}$
hybrid $M_{u\overline{u}g}$ do not exceed $1.56$ MeV.

We note that the spin-spin terms give an $<9\%$ correction for the light
flavors; we have neglected the L-S and tensor potential, which we suppose to
be small$^{[14,\text{ }19]}$. But an explicit calculation can be made, to
obtain a rigorous result; the study will be interesting, essentially to
remove the degeneracy between the states of the Table 1.

\section{Conclusion}

In this work we use a QCD-motivated potential (Coulombic plus Linear) to
estimate masses of both light and heavy hybrid mesons, in the context of a
constituent Quark-Gluon Model, taking into account the spin-spin interaction
effects for light hybrids. Our results are in a good agreement with the
other methods, like Lattice QCD, QCD Sum Rules, Bag Model, ... and with the
experimental candidates ($1^{-\text{ }+}$ at $1400$ , $1600$ and $2000$ $MeV$%
). We find also that $1^{-\text{ }+}$ hybrid mesons may exist in two
weakly-mixed modes: QE ($l_{q\bar{q}}=1,l_{g}=0$ and $S_{q\bar{q}}=0$) and
GE ($l_{q\bar{q}}=0,l_{g}=1$ and $S_{q\bar{q}}=1$), the later being much
heavier.

A more rigorous description of the hybrid meson can be given by improving
the potential, to obtain a classification of all the states in Table 1,
adding tensor terms and spin-orbit terms (which can generate a more
significative mixing between the two QE-GE modes$^{\left[ 8\right] }$).

At this time, it is more interesting to focus essentially on the $1^{-\text{ 
}+}$ hybrid states, taking in account the recent experimental observations
of the $1^{-\text{ }+}$ hybrid candidates: $\pi _{1}\left( 1400\right) ,\
\pi _{1}\left( 1600\right) .$

\paragraph{{\protect\Large Acknowledgments}\newline
}

We are grateful to O. P\`{e}ne \emph{(Laboratoire de Physique Th\'{e}orique,
Univ. of Paris-sud)} for extremely useful discussions. We would like to
thank the Abdus Salam International Center for Theoretical Physics, in
Trieste (Italy), for their Fellowship to visit the Center by the
Associateship scheme, where a part of this work was done.

\end{document}